\documentclass[prl,twocolumn,amsmath,amssymb,floatfix]{revtex4}
\usepackage{times}
\usepackage{graphicx}
\begin{document}

\preprint{APS/123-QED}

%\title{Interpolation between strong and weak coupling}
\title{Electrostatic interactions of charged bodies from the weak to
  the strong coupling regime}

\author{Marius M. Hatlo}
\author{Leo Lue}
\email{leo.lue@manchester.ac.uk}
\affiliation{
School of Chemical Engineering and Analytical Science \\
The University of Manchester \\
PO Box 88 \\
Sackville Street \\
Manchester M60 1QD \\
United Kingdom
}

\date{\today}

\begin{abstract}
A simple field theory approach is developed to model the properties of
charged, dielectric bodies and their associated counterions.  This
predictive theory is able to accurately describe the properties of
systems (as compared to computer simulation data) from the weak
coupling limit, where the Poisson-Boltzmann theory works well, through
to the strong coupling limit.  In particular, it is able to
quantitatively describe the attraction between like-charged plates and
the influence of image charge interactions.
%In addition, the theory remains accurate even in the presence of
%dielectric bodies.
%
%A theory is developed to interpolate between the strong and the weak
%coupling regime for two similarly charged surfaces in the presence of
%counterions. 
\end{abstract}

\maketitle

%%%%%%%%%%%%%%%%%%%%%%%%%%%%%%%%%%%%%%%%%%%%%%%%%%%%%%%%%%%%%%%%%%%%%%%%%%%%%%%%
%%%%%%%%%%%%%%%%%%%%%%%%%%%%%%%%%%%%%%%%%%%%%%%%%%%%%%%%%%%%%%%%%%%%%%%%%%%%%%%%
%\section{Introduction}

%Open with a statement about the importance of understanding the
%properties of system containing highly charged objects, such as
%colloidal and biological solutions.
Electrostatic interactions play a major role in determining the
structure and thermodynamics of many colloidal and biological
solutions, which typically contain charged macromolecular structures
with low dielectric interiors, such as DNA, charged micelles, or
membranes.  These charged structures are always surrounded by
neutralizing counterions, and, in many cases, the properties of the
system can be mainly attributed to properties of the counterions
\cite{Naji_etal_2005}.

When the electrostatic interactions are weak, their contributions to
the system properties are accurately described by the
Poisson-Boltzmann (PB) theory.  However, as these interactions
strengthen, the PB theory becomes less and less accurate.
Perturbation methods, such as the loop expansion can be used to
systematically improve the theory; however, the first loop correction
offers only a small improvement
\cite{Naji_etal_2005,Podgornik_1990,Attard_etal_1988a}, and
higher-order corrections are increasingly complicated to evaluate.

When the electrostatic interactions are extremely strong (e.g., when
the surface charge of the macromolecular structures or the valency of
the counterions is high), the Poisson-Boltzmann theory can yield {\em
  qualitatively} incorrect predictions.  For example, in this regime,
the counterions can generate attractive forces between similarly
charged objects \cite{Guldbrand_etal_1984}.  This phenomenon cannot be
explained by the PB theory, but has been observed in Monte Carlo
simulations \cite{Guldbrand_etal_1984} and in experiments (e.g.,
condensation of DNA molecules \cite{Bloomfield_1997}, bundle formation
of filamentous actin \cite{Tang_Janmey_1996}.
%and aggregation of colloidal particles) and with integral equation
%theories \cite{Kjellander_Marcelja_1984}.
%
%and field theoretic methods \cite{Moreira_Netz_2001,Naji_etal_2004}.
%
In this strong coupling regime, the counterions ``collapse'' on the
neutralizing charged surfaces to form a highly interacting 2D
structure that resembles a confined one-component plasma (OCP)
\cite{Gann_etal_1979,Shklovskii_1999b,Levin_2002}.
In these systems, the average distance $a_\perp$ between the ions is
much larger than the average distance $z$ between the ions and the
charged surface (i.e.\ $z\ll a_\perp$).  Consequently, a single
particle theory provides a good description of the system.  This leads
to the strong coupling (SC) expansion
\cite{Shklovskii_1999b,Moreira_Netz_2001}, which has been quite
successful
\cite{Moreira_Netz_2002,Kanduc_Podgornik_2007,Jho_etal_2008}.

%becomes exact in the limit when $\Xi\equiv l_B/\mu\rightarrow \infty$.

However, many systems are in a regime where both the PB theory and SC
expansion are inaccurate.
%
%To understand these systems, it is useful to consider the behavior of
%counterions near a low dielectric body with a uniform surface charge
%density $\Sigma$.
%
The behavior of these systems can be rationalized in terms of a
correlation hole \cite{Nordholm_1984} --- a region of size $\sigma$
around each counterion where it is unfavorable for other counterions
to be located.  At length scales greater than $\sigma$, the
counterions are weakly correlated, while at shorter length scales, the
counterions are strongly correlated but fairly ``isolated''
\cite{Moreira_Netz_2001}.
%Incorporating the correlation hole into a simple Debye-H\"uckel
%theory leads to results that are within 10\% of Monte Carlo
%simulations for the 3D OCP from weak to strong coupling
%\cite{Nordholm_1984,Levin_2002}.  The difference between the strong
%and weak coupling regimes can be understood by considering how the
%counterions behave near the charged surfaces.
In the weak coupling regime, the counterions form a diffuse 3D layer,
and the size of the correlation hole is approximately equal to the
Bjerrum length $l_B=\beta q^2/\epsilon$ (where $q$ is the counterion
charge), the distance at which two counterions interact with energy
$k_B T$. In the SC limit, the size of the correlation hole becomes
equal to the average (2D) distance between the ions
$a_\perp=2\mu\sqrt{2l_B/\mu}$, where $\mu=(2\pi\beta q\Sigma)^{-1}$ is
the Gouy-Chapman length (where $\Sigma$ is the surface charge
density), the distance at which the interaction between a counterion
and the charged surface equals $k_BT$.

Based on this observation, Weeks and coworkers \cite{Chen_Weeks_2006}
and Santangelo \cite{Santangelo_2006} developed approaches that split
the interaction between the ions at short and long range.  The
long-range interaction is treated within a mean field approximation,
and the short-range interactions with a more precise approach (e.g.,
computer simulation, liquid state theory, etc.).  With an appropriate
value for $\sigma$, these approaches can successfully describe Monte
Carlo results for the full range of electrostatic coupling.  However,
the value of $\sigma$ is determined empirically.  Additionally, these
approaches are not capable of describing systems with dielectric
inhomogeneities.

In this work, we present a self-consistent theory that is in good
agreement with Monte Carlo simulations at weak, intermediate, and
strong coupling.  The theory is similar in idea to the work of Weeks
and coworkers \cite{Chen_Weeks_2006} and Santangelo
\cite{Santangelo_2006}, however, the parameter $\sigma$ is calculated
consistently from the partition function, rather than chosen
empirically or adjusted to fit data.  In addition, the theory
accurately describes the presence of dielectric bodies, even in the SC
limit, which has not been demonstrated by any previous theory.
For the two plate system in the presence of image charges, the system
undergoes a transition from from a two peak density profile to a one
peak density profile.

%This effect has not been successfully is not addressed by these works
%and is not captured successfully by either the PB or the SC theories.

%between the counterion containing medium and the macroscopic charged
%object

%%%%%%%%%%%%%%%%%%%%%%%%%%%%%%%%%%%%%%%%%%%%%%%%%%%%%%%%%%%%%%%%%%%%%%%%%%%%%%%%
%%%%%%%%%%%%%%%%%%%%%%%%%%%%%%%%%%%%%%%%%%%%%%%%%%%%%%%%%%%%%%%%%%%%%%%%%%%%%%%%
%\section{Theory}

We limit our attention to systems composed of a fixed charge
distribution $\Sigma({\bf r})$ that is surrounded by a neutralizing
cloud of counterions, which are point charges of magnitude $q$
immersed in a medium with dielectric constant $\epsilon$ and possibly
in the presence of dielectric inhomogeneities.  The total
electrostatic energy $H$ of the system is given by
\begin{equation}
\label{eq:H_elec}
\begin{split}
H &= \frac{1}{2}\int d{\bf r}d{\bf r'}
      Q({\bf r})G_0({\bf r},{\bf r'})Q({\bf r'})
\\ & \qquad
- \sum_{k}\frac{q^2}{2}G_{\rm free}({\bf r}_k,{\bf r}_k)
%- \sum_{k} e^{\rm se}({\bf r}_{k})
\end{split}
\end{equation}
where $Q({\bf r})=q\sum_k\delta^d({\bf r}-{\bf r}_k)+\Sigma({\bf r})$
is the total charge density, ${\bf r}_k$ is the position of counterion
$k$, $G_0$ is the Green's function of the associated electrostatics
problem (including the effects of dielectric objects), and $G_{\rm
  free}$ is the Green's function in the absence of dielectric
inhomogeneities.
%The self energy of the point charges is
%\begin{equation}
%e^{\rm se} = \frac{q^2}{2}G_{\rm free}({\bf r},{\bf r}).
%\end{equation}
%where $G_{\rm free}$ is the Green's function in the absence of
%dielectric inhomogeneities.

To separate short and long wavelength phenomena, we split
\cite{Chen_Weeks_2006,Rodgers_etal_2006,Santangelo_2006} the Green's
function $G_0$ into a short wavelength $G_s$ and a long wavelength
$G_l$ component
\begin{equation}
G_0({\bf r},{\bf r'}) = G_s({\bf r},{\bf r'})+G_l({\bf r},{\bf r'})
%= \frac{e^{-|{\bf r}-{\bf r'}|/\sigma}}{|{\bf r}-{\bf r'}|}
%- \frac{e^{-|{\bf r}-{\bf r'}|/\sigma}-1}{|{\bf r}-{\bf r'}|}
\end{equation}
where $G_s=(1-{\mathcal P})G_0$, and $G_l={\mathcal P}G_0$.  The
operator $\mathcal{P}$ filters out the short wavelengths; its specific
form is arbitrary, and in this work we use ${\mathcal P} =
[1-\sigma^2\nabla^2+\sigma^4\nabla^4]^{-1}$, 
%which is simlar to the form used by Weeks and coworkers
%\cite{Chen_Weeks_2006},
where $\sigma$ is the length scale which divides the long from the
short wavelength phenomena.  
%This choice of ${\mathcal P}$ allows the final equations to be cast
%solely in terms of differential equations, rather than involving
%integral equations.  which resembles the Gaussian projection operator
%used by Chen and Weeks \cite{Chen_Weeks_2006}, however with a
%$\bar{\sigma}=2\sigma$.  The stationary value of $\sigma$ for
%$\Xi=100$ is $\sigma = 7.1$, which is very close to the best fit
%value $\bar{\sigma}=15, (\sigma=7.5)$ obtained by Chen and Weeks
%\cite{Chen_Weeks_2006}.  Other choices of the projection operator are
%possible (e.g., $\exp(-\sigma^2\nabla^2/4)$, see
%Ref.~\cite{Chen_Weeks_2006}).

Equation~(\ref{eq:H_elec}) can be written as:
\begin{equation}
\begin{split}
H &= %\frac{\lambda_l}{2}\int d{\bf r}d{\bf r'}
\frac{1}{2}\int d{\bf r}d{\bf r'}
      Q({\bf r}){G}_l({\bf r},{\bf r'})Q({\bf r'})
  + E^{\rm se}
\\ & \quad
%+ \frac{q^2\lambda_s}{2}\sum_{jk}{G}_s({\bf r}_j,{\bf r}_k)
+ \frac{q^2}{2}\sum_{jk}{G}_s({\bf r}_j,{\bf r}_k)
\\ & \quad
  + \sum_{k}\left[
    u({\bf r}_k)
%    - \frac{q^2\lambda_s}{2} {G}_s({\bf r}_k,{\bf r}_k)
%    - \frac{q^2\lambda_l}{2}{\mathcal P}\delta G_0({\bf r}_k,{\bf r}_k)
    - \frac{q^2}{2} {G}_s({\bf r}_k,{\bf r}_k)
    - \frac{q^2}{2}{\mathcal P}\delta G_0({\bf r}_k,{\bf r}_k)
  \right]
\end{split}
\end{equation}
%where $\lambda_s$ and $\lambda_l$ (which are equal to unity) are used
%to account for the order of the approximation.
where $\delta G_0=G_0-G_{\rm free}$, $u({\bf r})$ is a one-particle
interaction potential given by
\begin{equation}
\begin{split}
\label{eq:u-1}
u({\bf r}) &= q \int d{\bf r}' G_s({\bf r},{\bf r}')\Sigma({\bf r}')
+ \frac{q^2}{2}\delta G_0({\bf r},{\bf r})
\\ & \qquad
- \frac{q^2}{2}\mathcal{P}G_{\rm free}({\bf r},{\bf r}),
\end{split}
\end{equation}
and $E^{\rm se}$ is the self energy of the fixed charges, defined as
\begin{equation}
\label{eq:self}
E^{\rm se} = \frac{1}{2}\int d{\bf r}d{\bf r'}
  \Sigma({\bf r})G_s({\bf r},{\bf r'})\Sigma({\bf r'}).
\end{equation}

By performing a Hubbard-Stratonovich transformation
\cite{Hubbard_1959,Stratonovich_1957} twice on the grand partition
function of the system, two fields are introduced: $\psi_l$, which is
associated with $G_l$ and represents interactions at length scales
greater than $\sigma$, and $\psi_s$, which is associated with $G_s$
and represent interactions at length scales less than $\sigma$.
In the approximation scheme we pursue, the one-particle contribution
to the partition function is treated exactly, while the interaction
between the particles is treated approximately.
The field $\psi_s$ is strongly fluctuating and coupled not only to
itself, but also to the field $\psi_l$.  To evaluate the functional
integration over $\psi_s$, we use a cumulant expansion, truncated at
first order.  This leads to an approximation similar to the SC
expansion of Moreira and Netz
\cite{Moreira_Netz_2000a,Moreira_Netz_2001}.

Performing the functional integration over $\psi_s$, we get an
effective field theory for the long wavelength system.  The expression
for the grand partition function $Z_{\rm G}$ is
\begin{equation}
\label{eq:H_l}
\begin{split}
\ln Z_{\rm G}[\gamma,\Sigma] 
&= - \beta E^{\rm se}
+ \ln \left\{
  \frac{1}{\mathcal{N}_{l}}\int \mathcal{D}\psi_l(\cdot) e^{-H_l[\psi_l]} 
  \right\}
\end{split}
\end{equation}
and $H_l$ is the effective Hamiltonian, which is a functional of the
field $\psi_l$ given by 
\begin{equation}
\begin{split}
- H_l[\psi_l] &\approx %-\frac{1}{2\beta\lambda_l}
-\frac{1}{2\beta}
  \int d{\bf r}d{\bf r'}
     \psi_l({\bf r})G^{-1}_l({\bf r},{\bf r}')\psi_l({\bf r}')
\\ & \quad  
+ \int d{\bf r}\Lambda^{-3} e^{\gamma-qi\psi_l({\bf r})
  - \beta u({\bf r}) 
%  + \frac{\beta q^2\lambda_l}{2}\delta {G}_l({\bf r},{\bf r}) }
  + \frac{\beta q^2}{2}\delta {G}_l({\bf r},{\bf r}) }
\\ & \quad  
- \int d{\bf r}\Sigma({\bf r}) i{\psi}_l({\bf r})
%+ \mathcal{O}((\Xi\lambda_s)^{-1})
.
\end{split}
\end{equation}
%Here the Green's function is scaled by the Gouy-Chapman length
%($\mu\tilde{G}_l=G_l$).
%
%We can identify the parameter $\lambda_l=\mu/\mu_{\rm eff}$, where
%$\mu_{\rm eff}$ is an effective Gouy-Chapman length, accounting for
%how the strongly coupled ions effectively reduce the surface charge of
%the plates \cite{Lau_Pincus_2002}.  
%

%as its the correlations between the particles that determines if we
%are in the strong coupling regime or the weak coupling regime, not the
%interaction of the particles with its image or the surface charge.

%The idea is that $q^2\lambda_l$ is
%representative of the average interaction between the particles at
%long wavelengths, which is assumed small while $q^2\lambda_s$
%represent the magnitude of the average interaction at short
%wavelengths assumed large.  
%The long wavelength fluctuations are
%determined by $\Xi\lambda_l$, while the short range fluctuations are
%determined by $\Xi \lambda_s$.  To obtain this exact one-particle
%contribution the interaction of one particle with its image has been
%added and subtracted.

The long-wavelength field $\psi_l$ is weakly fluctuating, so a
mean-field approximation is sufficient to evaluate the functional
integral over $\psi_l$ in Eq.~(\ref{eq:H_l}):
%(i.e.\ keeping terms to $\mathcal{O}(\Xi\lambda_l$))
\begin{equation}
\begin{split}
\ln Z_{\rm G}[\gamma,\Sigma]  &= 
+ \frac{1}{2\beta}\int d{\bf r}d{\bf r'}
  i\bar{\psi}_l({\bf r})G_l^{-1}({\bf r},{\bf r'})i\bar{\psi}_l({\bf r})
\\ & \quad 
- \int d{\bf r}\Sigma({\bf r}) i\bar{\psi}_l({\bf r})
- \beta E^{\rm se}
\\ & \quad  
+ \int d{\bf r}\Lambda^{-3} e^{\gamma-qi\bar{\psi}_l({\bf r})-\beta u({\bf r})} 
%+ \mathcal{O}((\Xi\lambda_s)^{-1},\Xi\lambda_l),
,
\end{split}
\end{equation}

The value of the mean field $\bar{\psi}_l({\bf r})$ is determined by
solving the Poisson equation
\begin{equation}
\label{eq:poisson}
% \frac{1}{\beta}\hat{G}^{-1}_l i\bar{\psi}_l({\bf r})
%= \frac{1}{\beta}\hat{G}^{-1}_0{\mathcal P}^{-1} i\bar{\psi}_l({\bf r})
- \frac{1}{4\pi}\nabla\cdot\epsilon({\bf r})\nabla \phi({\bf r})
= \Sigma({\bf r}) + q\rho({\bf r}),
\end{equation}
where the electric potential is defined as $\beta\phi={\mathcal
  P}^{-1}i\bar{\psi}_l$, and the counterion density distribution
$\rho({\bf r})$ is given by
\begin{equation}
\label{eq:density}
\rho({\bf r})= \Lambda^{-3}
e^{\gamma   - qi\bar{\psi}_l({\bf r}) - \beta u({\bf r})}.
\end{equation}
Note that the density depends on the field $\bar{\psi}_l={\mathcal
  P}\phi$ rather than on the electric potential $\phi$, as in the PB
theory.
In the systems we consider, where there are only counterions, the
chemical potential $\gamma$ is determined by the electroneutrality
constraint.
%Because the number of counterions is fixed by the
%electroneutrality condition, it is more convenient to work with the
%Helmholtz free energy $F$, which is given by
%\begin{equation}
%\begin{split}
%\label{eq:free-energy}
%\beta F[\rho] &= 
%\int d{\bf r}\rho({\bf r})[\ln\rho({\bf r})-1]
%\\ & \quad
%+ \frac{1}{8\pi} \int d{\bf r} \epsilon({\bf r}) 
%  \nabla\phi({\bf r})\cdot\nabla i\bar{\psi}_l({\bf r})
%\\ & \quad
%+ \int d{\bf r} \rho({\bf r}) \beta u({\bf r}) + \beta E^{\rm se}.
%\end{split}
%\end{equation}
%The first term is related to the ``entropic'' contribution of the
%counterions.  The second term is the energy of the electrostatic
%field.  The third term is the contribution of the one-body
%interactions of the counterions.  The final term is the
%short-wavelength contribution to the self energy of the fixed charges.

%The free energy should be independent on the parameter $\sigma$;
%however, 
All properties of the system should be independent of the parameter
$\sigma$; however, because the theory is approximate, there will be a
dependence on the value of $\sigma$.  Based on this, we determine the
value of $\sigma$ by requiring that the
%free energy is stationary with respect to this 
grand partition function is stationary with respect to $\sigma$
(i.e.\ $\partial \ln Z_G[\gamma]/\partial\sigma=0$).  This is similar
to the optimized random phase approximation
\cite{Andersen_Chandler_1972}.

%%%%%%%%%%%%%%%%%%%%%%%%%%%%%%%%%%%%%%%%%%%%%%%%%%%%%%%%%%%%%%%%%%%%%%%%%%%%%%%%
%%%%%%%%%%%%%%%%%%%%%%%%%%%%%%%%%%%%%%%%%%%%%%%%%%%%%%%%%%%%%%%%%%%%%%%%%%%%%%%%
%\section{One plate}

Now we apply the theory to a system of counterions confined to one
side of a plate with dielectric constant $\epsilon'$ and a uniform
surface charge $\Sigma({\bf r})=\delta(z)\Sigma$, where $z$ is the
distance from the surface of the plate.  For this system, the one-body
potential of the counterions (see Eq.~(\ref{eq:u-1})) reduces to
\begin{equation}
\begin{split}
\beta u({\bf r})
&= -\frac{2\sigma}{\sqrt{3}\mu}(1+\Delta)e^{-\frac{\sqrt{3}z}{2\sigma}}
\cos\frac{z}{2\sigma}
%\\ & \qquad
+ \frac{l_B\Delta}{4z} -\frac{l_B}{2\sqrt{3}\sigma},
\end{split}
\end{equation}
where $z$ is the distance from the surface of the plate, and
$\Delta=(\epsilon'-\epsilon)/(\epsilon'+\epsilon)$.  The self energy
of the surface charge (see Eq.~(\ref{eq:self})) is $\beta E^{\rm se}/N
= \frac{\sigma}{\sqrt{3}\mu}(1+\Delta)$ where $N$ is the total number
of counterions in the system.  The strength of the electrostatic
interactions is characterized by the parameter $\Xi\equiv l_B/\mu$.

\begin{figure}[tb]
\begin{center}
\includegraphics[clip,width=\columnwidth]{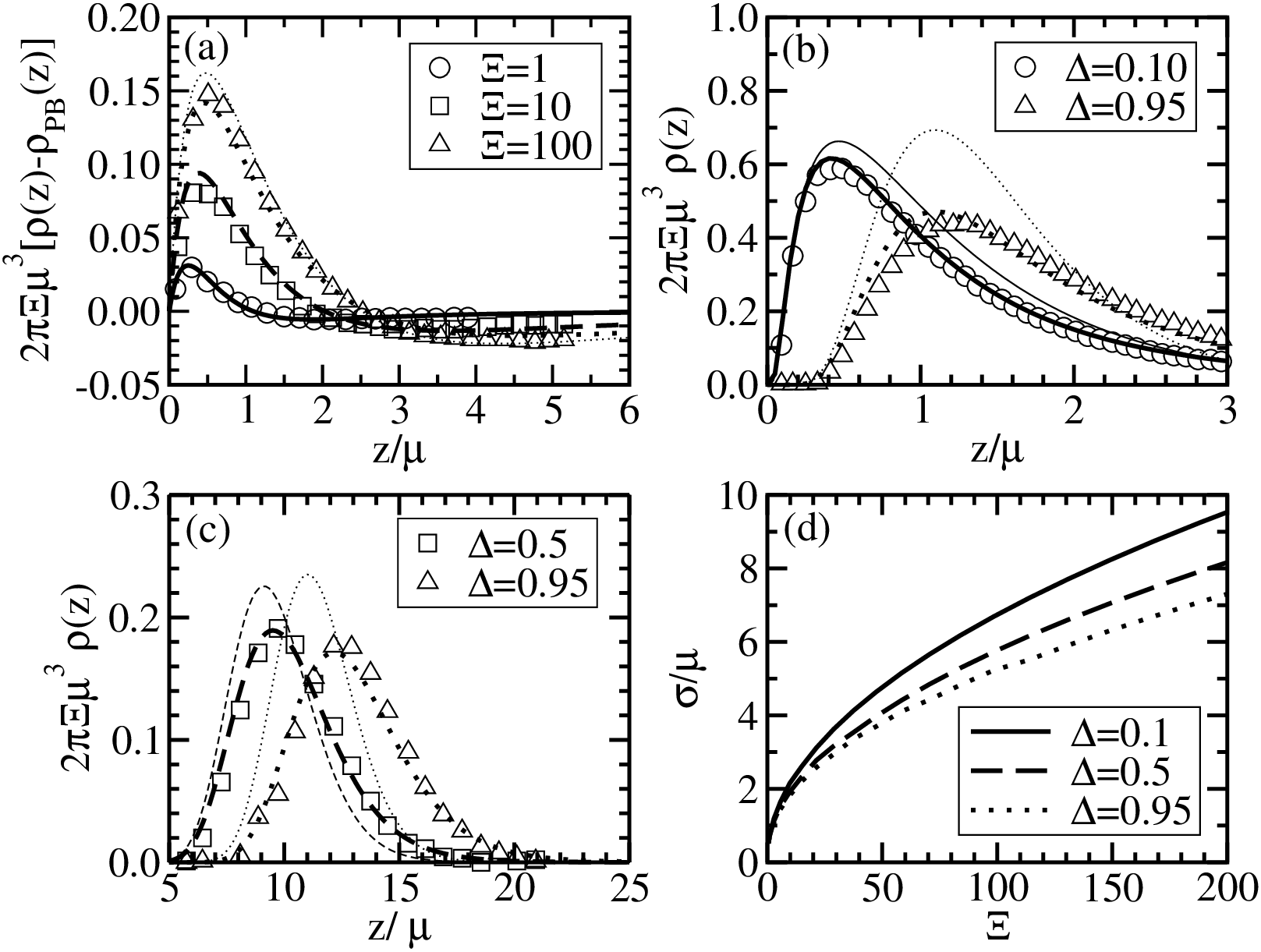}
\caption{\label{fig:density-1plate} 
(a) Counterion density profile near a single charged plate with
  $\Delta=0$.
%for (i) $\Xi=1$ (solid lines, circles), (ii) $\Xi=10$ (dashed lines,
%squares), and (iii) $\Xi=100$ (dotted lines, triangles).
%
Counterion density profile for (b) $\Xi=10$ and (c) $\Xi=1000$ near a
single charged, dielectric plate.
%with (i) $\Delta=0.1$ (solid lines, circles), (ii) $\Delta=0.5$
%(dashed lines, squares), and $\Delta=0.95$ (dotted lines, triangles).
%
(d) Dependence of the parameter $\sigma$ with the coupling parameter $\Xi$.
%for (i) $\Delta=0.1$ (solid line), (ii) $\Delta=0.5$ (dashed line),
%and (iii) $\Delta=0.95$ (dotted line).
%
The symbols are Monte Carlo simulation data
\cite{Moreira_Netz_2002,Moreira_Netz_2002b,moreira_2001}, the thin
lines are the predictions of the SC expansion
\cite{Moreira_Netz_2002,Moreira_Netz_2002b,moreira_2001}, and the
thick lines are from the present work.
}
\end{center}
\end{figure}

%Predictions for the counterion density profile for the single plate
%system are plotted in Fig.~\ref{fig:density-1plate}.
%
The deviations of the counterion density profile from the predictions
of the PB theory for the single plate system are plotted in
Fig.~\ref{fig:density-1plate}a in the case when there is no dielectric
interface (i.e.\ $\Delta=0$).  As the strength of the electrostatic
interactions increases, the PB theory overpredicts the repulsion
between the counterions and the ion cloud associated with the charged surface, and,
%due to its neglect of correlation effects, and,
consequently, it underestimates the adsorption of the counterions to
the surface.  The SC limit (given by the thin line) is only approached
at fairly high values of the coupling parameter ($\Xi>100$).  The
predictions of the present theory are, however, in good agreement with
Monte Carlo simulation data, exhibiting the crossover of the density
profile from the PB theory to the SC limit with increasing values of
$\Xi$.

%Monte Carlo simulations indicate \cite{Naji_etal_2005} that the ions
%obey the strong coupling theory when $z\ll a_\perp$, and a mean field
%theory (e.g., PB theory) when $z\gg a_\perp$.  For the pressure, much
%higher values of the coupling parameter are required before the
%strong coupling limit is approached.

%%%
%%%
%%%
%%%

%For the case of counterions confined between two uniformly charged
%plates separated by a distance $d$, the variation of $\sigma$ with
%plate spacing and coupling strength is shown in
%Fig.~\ref{fig:sigma-interpolation}(b).  The correlation length
%$\sigma$ decreases with the plate spacing, which is due to the fact
%that the average distance between the counterions become smaller as
%the ion atmospheres of the two plates start to overlap.

When the plate has a low dielectric interior
(i.e.\ $\epsilon'<\epsilon$, $\Delta>0$), repulsive image charge
interactions with the plate repel the counterions from the surface,
which oppose the attractive interactions with the surface charge.
Consequently, the counterions no longer collapse onto the surface, but
instead peak at a distance away from the surface
\cite{Henderson_etal_2005,Bhuiyan_etal_2007,Hatlo_etal_2008}.
%
%This system has been extensively studied in the weak coupling regime,
%with results that compare well with Monte Carlo simulations
%\cite{Hatlo_etal_2008,Bhuiyan_etal_2007,Henderson_etal_2005} and in
%the strong coupling limit \cite{Moreira_Netz_2002}.
%
Counterion density profiles for this situation are plotted in
Figs.~\ref{fig:density-1plate}b and c.  When $\Delta>0$, the SC
expansion fails to accurately describe the Monte Carlo data, even well
into strong coupling regime ($\Xi=1000$).  The failure of the SC
theory is due to the fact that the average distance between the ions
and the distance to the charged plate are the same order of magnitude.
Interactions between the counterions become significant, and the
physical basis for the SC is no longer fulfilled.

Our approach is able to overcome these difficulties because in
addition to including the one-body interactions with the plate
interactions, which occur at length scales less than $\sigma$, it also
accounts for the interaction between ions at length scales greater
than $\sigma$.  The variation of $\sigma$ with the coupling parameter
$\Xi$ is plotted for various values of $\Delta$ in
Fig.~\ref{fig:density-1plate}d.  For high values of the coupling
parameter, the size of the correlation hole quickly approaches a
constant with respect to the spacing $a_{\perp}$ between the
counterions.  As $\Delta$ increases, the size of the correlation hole
decreases, reflecting the increasing importance of the interactions
between the counterions.
%
%
%This reflects the larger volume that the counterions can occupy, which
%translates into a greater spacing.

%%%%%%%%%%%%%%%%%%%%%%%%%%%%%%%%%%%%%%%%%%%%%%%%%%%%%%%%%%%%%%%%%%%%%%%%%%%%%%%%
%%%%%%%%%%%%%%%%%%%%%%%%%%%%%%%%%%%%%%%%%%%%%%%%%%%%%%%%%%%%%%%%%%%%%%%%%%%%%%%%
%\section{Two plate}

\begin{figure}[tb]
\begin{center}
\includegraphics[clip,width=\columnwidth]{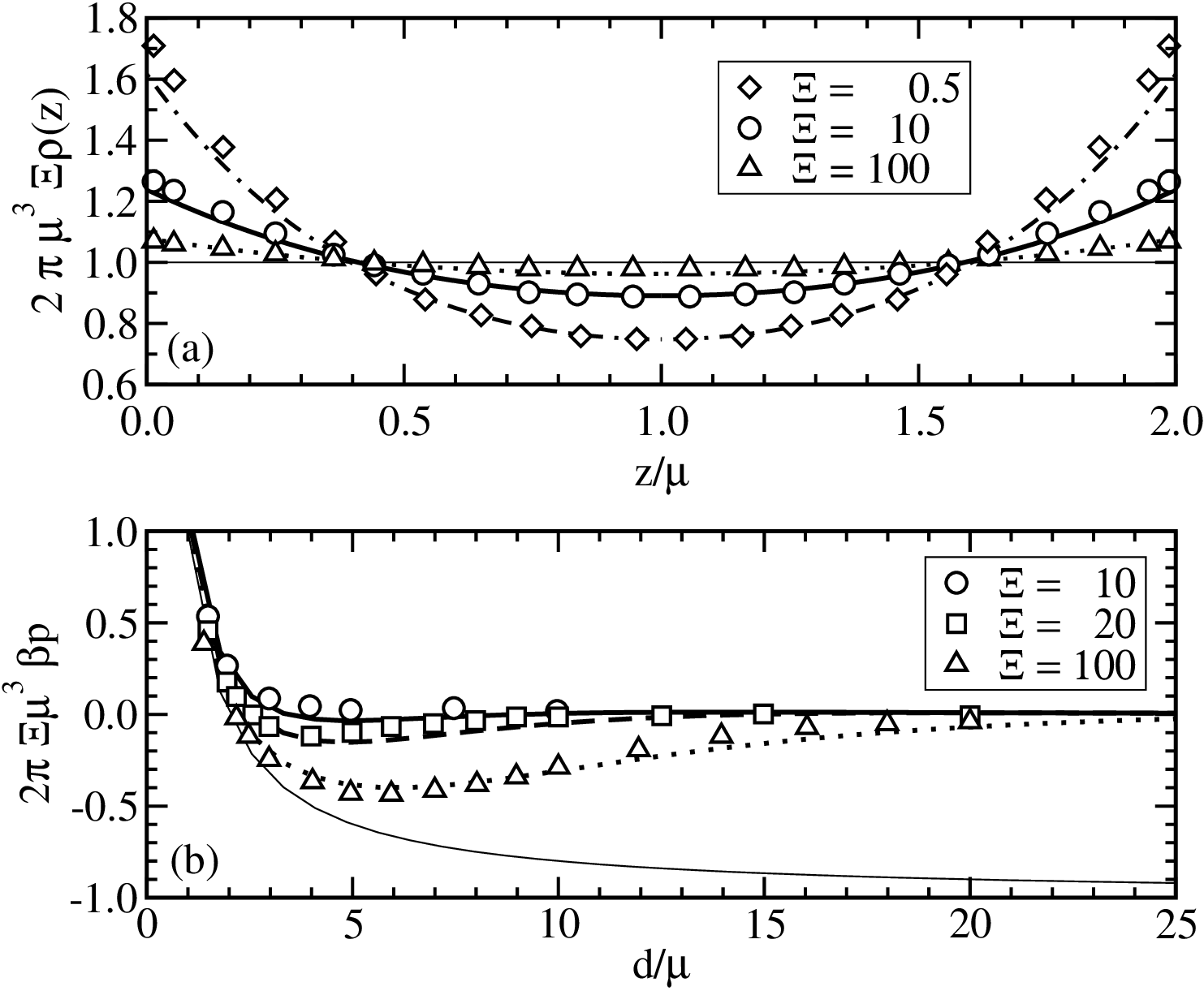}
\caption{\label{fig:twoplate} 
(a) Density profile for counterions confined between two charged
  plates with $\Delta=0$ separated by distance $d=2\mu$.  (b) Force
  between the plates as a function of their separation $d$.
%
%(i) $\Xi=0.5$ (dashed-dotted lines, diamonds), 
%(ii) $\Xi=10$ (solid lines, circles), 
%(iii) $\Xi=20$ (dashed lines, squares), and 
%(iv) $\Xi=100$ (dotted lines, triangles).
%
The symbols are simulations data \cite{Moreira_Netz_2001}, the thin
lines are the prediction of the SC theory \cite{Moreira_Netz_2001},
and the thick lines are from this work.  }
\end{center}
\end{figure}

Now we consider systems where the counterions are confined between two
uniformly charged plates, separated by a distance $d$.
Fig.~\ref{fig:twoplate}a shows the counterion density profile where
the plates have dielectric constant $\epsilon'=\epsilon$ and are
separated by a distance $d/\mu=2$; results for the pressure are shown
in Fig.~\ref{fig:twoplate}b.  At weak couplings, the force between the
plates is strictly repulsive, but as $\Xi$ increases, a region of
attraction develops at intermediate plate separations.  The
predictions of this work are in quantitative agreement with the
computer simulation data \cite{Moreira_Netz_2001}, while the SC limit
is applicable only at extremely high values of the coupling parameters
($\Xi\sim10^5$ for the pressure, see Ref.~\cite{Moreira_Netz_2001}).
%
%To put these results in perspective, we note that for silica plates
%with a bare charge density of approximately $\Sigma=-0.02$~$e$/\AA$^2$
%\cite{Trulsson_etal_2006} (where $e$ is the elementary charge), the
%coupling parameter is $\Xi \approx 6|q/e|^3$ in water at room
%temperature.  This implies that for trivalent counterions, $\Xi\approx
%160$, which would be well into the attractive regime, in agreement
%with experimental measurements \cite{Zohar_etal_2006}.

%%%
%%%
%%%
%%%

\begin{figure}[t]
\begin{center}
\includegraphics[clip,width=\columnwidth]{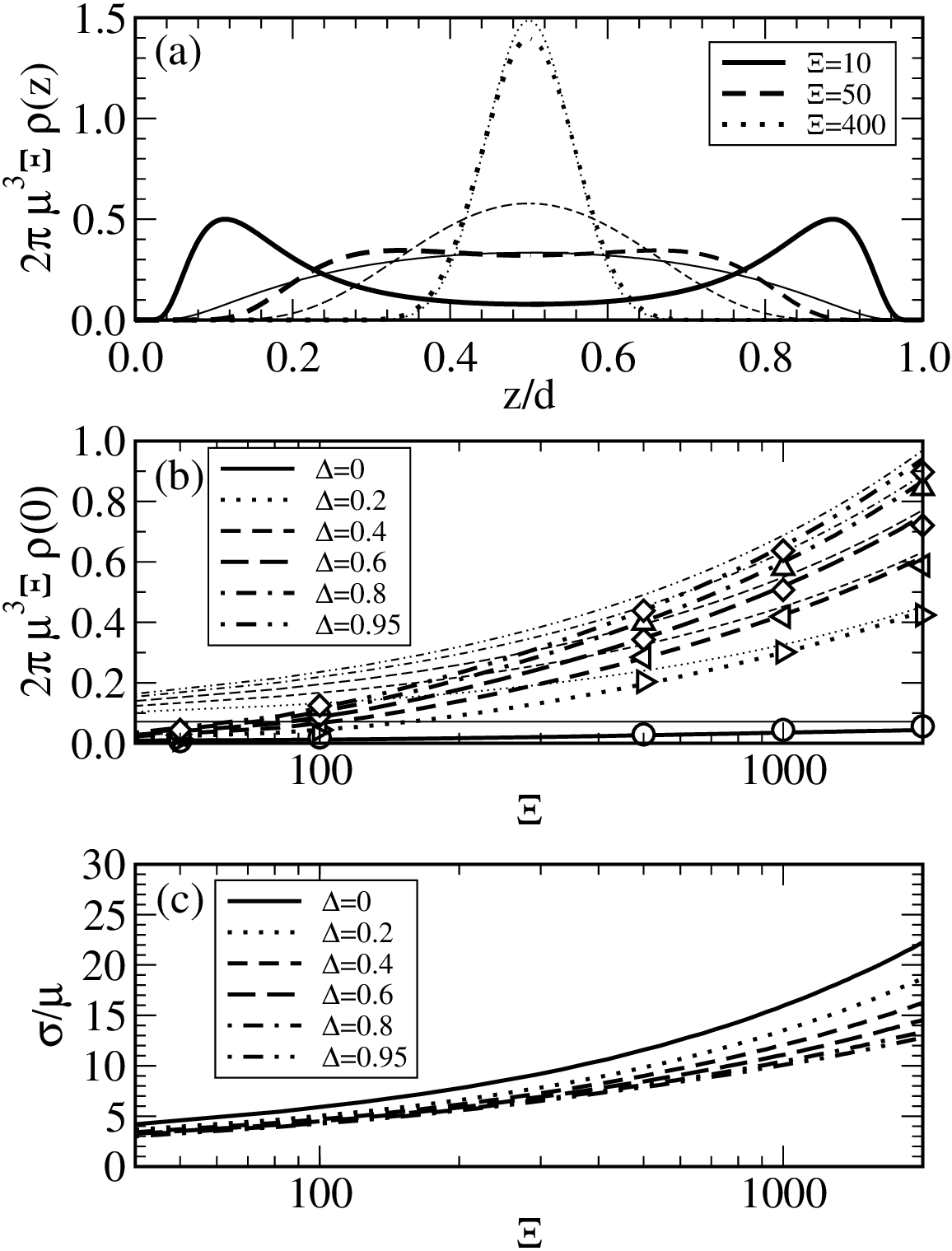}
\caption{\label{fig:2-plate-image} 
(a) Density profile for counterions confined between charged plates with
$\Delta=0.98$ and $d=10\mu$.
% 
%(i) $\Xi=10$ (solid lines), 
%(ii) $\Xi=50$ (dashed lines), and 
%(iii) $\Xi=400$ (dotted lines).
%
(b) Counterion density at the mid-point between charged plates
  separated by distance $d=14\mu$ and (c) the corresponding variation
  of the parameter $\sigma$.
The symbols are Monte Carlo simulation data \cite{Jho_etal_2008}, the
thin lines are the prediction of the SC theory
\cite{Moreira_Netz_2001,Kanduc_Podgornik_2007,Jho_etal_2008}, and the
thick lines are from this work.  }
\end{center}
\end{figure}

Decreasing the dielectric constant of the plates
($\epsilon'<\epsilon$) leads
\cite{Kanduc_Podgornik_2007,Hatlo_Lue_2008} to a qualitatively
different counterion density profile, as shown in
Fig.~\ref{fig:2-plate-image}a.  
For sufficiently large plate separations, there is a counterion peak
next to each plate.
When the distance between the plates is smaller or comparable to the
average distance between the ions ($a_\perp/\mu\approx\sqrt{2\Xi}$, so
$d/\mu<\sqrt{2\Xi}$), the repulsive image charge interactions push
the counterions into a single peak in the middle of the plates.  At
these separations, the distance between an ion and its image charge is
comparable to the average distance between the ions.
The variation of the mid-point counterion density with the coupling
constant is shown in Fig.~\ref{fig:2-plate-image}b.  The predictions
of the present work are in good agreement recent Monte Carlo
simulation results for all conditions examined.  The SC theory
\cite{Kanduc_Podgornik_2007,Jho_etal_2008} is only accurate when
$d/\mu\ll\sqrt{2\Xi}$.
%The pressure can be obtained by taking the derivative of the free energy.
%It can be demonstrated that in the absence of a dielectric jump, the
%theory obeys the contact value theorem.
%
The parameter $\sigma$ (see Fig.~\ref{fig:2-plate-image}c) increases
with increasing values of $\Xi$ and decreasing values of $\Delta$, in
a similar manner with the one plate case.

%%%
%%%
%%%
%%%

The approach developed here is applicable to general geometries (e.g.,
spherical or cylindrical) in the same manner as the Poisson-Boltzmann
equation.  The only difference will be the form of the fixed charge
density $\Sigma({\bf r})$.  Once this is given, the one particle
potential is obtained by Eq.~(\ref{eq:u-1}) and the density by
Eq.~(\ref{eq:density}).  The electric potential is obtained from the
solution of the Poisson equation (see Eq.~(\ref{eq:poisson})), which
can be solved by standard methods
\cite{Chen_Weeks_2006,Santangelo_2006}.
%For example, this approach yields very accurate results for
%the 3D OCP, where the fixed charge distribution is a uniform
%background.  In addition, the theory works well even when co-ions
%(i.e., salts) are present.
%
This theory can also be systematically improved by either increasing
the order of the cumulant expansion for the integration over $\psi_s$
or by going beyond the mean field theory
%(e.g., loop
%expansion or variational perturbation approximation) 
for $\psi_l$.

We thank Prof.~C.~Holm and Dr.~M.~Sega for providing us with
unpublished simulation data for the two plate system.  MM Hatlo
acknowledges support from an EC Marie Curie Fellowship
(MEST-CT-2004-503750).

%%%%%%%%%%%%%%%%%%%%%%%%%%%%%%%%%%%%%%%%%%%%%%%%%%%%%%%%%%%%%%%%%%%%%%%%%%%%%%%%
%%%%%%%%%%%%%%%%%%%%%%%%%%%%%%%%%%%%%%%%%%%%%%%%%%%%%%%%%%%%%%%%%%%%%%%%%%%%%%%%
%\bibliography{strong} 
%\bibliographystyle{apsrev}

%%%%%%%%%%%%%%%%%%%%%%%%%%%%%%%%%%%%%%%%%%%%%%%%%%%%%%%%%%%%%%%%%%%%%%%%%%%%%%%%
%%%%%%%%%%%%%%%%%%%%%%%%%%%%%%%%%%%%%%%%%%%%%%%%%%%%%%%%%%%%%%%%%%%%%%%%%%%%%%%%
%%%%%%%%%%%%%%%%%%%%%%%%%%%%%%%%%%%%%%%%%%%%%%%%%%%%%%%%%%%%%%%%%%%%%%%%%%%%%%%%
\end{document}